# The First Turbulence

Carl H. Gibson

Departments of Mechanical and Aerospace Engineering

and Scripps Institution of Oceanography,

Center for Astrophysics and Space Sciences,

University of California at San Diego

La Jolla, CA 92093-0411

cgibson@ucsd.edu, http://www-acs.ucsd.edu/~ir118

**Abstract**

Chaotic, eddy-like motions dominated by inertial-vortex forces cascade from Planck scales to larger strong-force-freeze-out scales in a turbulent hot-big-bang cosmological model with inflation. A nonlinear mechanism is proposed for the initial quantum-gravitational-dynamics epoch that produces space-time-energy and the first large Reynolds number turbulence and turbulent mixing. The resulting Batchelor-Obukhov-Corrsin temperature spectrum is preserved as the first fossil-temperature-turbulence by the exponential inflation of space stretching the temperature fluctuations beyond the horizon scale $ct$ of causal connection faster than light speed $c$ in time $t$. Stretching continues as the universe expands, but now all scales of big-bang fossil turbulence are within our horizon. The expected temperature gradient spectrum increases toward a maximum at the smallest (fossilized Planck) scales where big-bang turbulence originates, contrary to models that explain the cosmic microwave background temperature spectrum as acoustic oscillations in the primordial plasma superposed on a flat or drooping background spectrum. The hot radiation-dominated plasma epoch was too viscous for sound waves to persist with gluon and photon viscous damping. In the present model, fossil-temperature-turbulence from the big bang reentered the horizon after inflation and triggered both the nucleosynthesis of light elements and the viscous-gravitational formation of structure in the plasma. Buoyancy and viscous forces inhibited turbulence in the plasma, as indicated by observations that T/T in the cosmic microwave background radiation is only $10^{-5}$ with a smooth wavenumber $k^{1/6}$ spectral form except at a single peak. This spectral peak is due to the first gravitational structure formation, not sound, and the rest of the spectrum is fossil big-bang turbulence.

## 1. Introduction

Turbulent motions, turbulent mixing, and turbulent diffusion are crucial to the evolution of natural fluids. However, because the conservation equations for turbulence are nonlinear and because no systematic method exists for solving nonlinear equations, turbulence remains as an unsolved problem of classical physics. A variety of definitions of turbulence may be found in the literature. Even the direction of the turbulence cascade is controversial. The plan of the present paper is to propose a universal definition of turbulence and a universal direction of the turbulent cascade, and then identify from theory and observations the time when turbulence first occurred as the universe evolved. The proposed definition is narrow, and is based on a nonlinear term in the momentum conservation equation called the inertial-vortex force that persists at all scales of the self-similar turbulence cascade. By this definition, turbulence always starts at small scales and cascades to larger scales until constrained by boundaries, other forces, or by the scale of causal connection. The proposed quantum gravitational instability mechanism at Planck scales and the expansion of the hot-big-bang universe before the strong-force-freeze out time matches the definition, and is thus the first turbulence.

A distinctive property of turbulence is the irreversibility of its effects. Linear waves come and go without leaving a trace, but turbulence and turbulent mixing produce unique, persistent, fingerprints termed fossil turbulence from which previous hydrophysical parameters like dissipation rates, viscosities and diffusivities may be extracted by the process of hydropaleontology, Gibson 1999. Information preserved by fossil turbulence is valuable in interpreting oceanographic, atmospheric and astrophysical measurements where little or no possibility exists to detect the turbulence in its active state. Cosmological information about the first turbulence and the first gravitational structure formation is preserved by the cosmic microwave background temperature spectrum, Gibson 2000.

## 2. Definition of turbulence

Webster's dictionary defines a turbulent flow as one in which the velocity at a given point varies erratically in magnitude and direction. This definition is perhaps the one most commonly used in the fields of fluid mechanics, magnetohydrodynamics, oceanography, and atmospheric sciences. However, many counterexamples to such a broad definition of turbulence are immediately apparent. Everything that wiggles is not turbulence. Viscous sublayers have random velocities because their boundary conditions are random, and are easily recognized as nonturbulent flows because their Reynolds numbers are subcritical. Oceanic and atmospheric flows always have a random component, but are generally dominated by buoyancy or Coriolis forces at large scales and are decomposed into classes of nonturbulent linear wave motions because their Froude and Rossby numbers are subcritical at large scales even though Reynolds numbers are supercritical.

Several oscillations in magnetohydrodynamic flows are chaotic and called turbulent that might be considered nonturbulent in other branches of fluid mechanics. Clearly, we need a definition of turbulence that is length scale dependent, and which recognizes that only one nonlinear force must dominate all other forces within the turbulent range of scales. No other definition can permit the universal similarity hypothesized by Kolmogorov based on length scale $L_K = (\nu^3/\varepsilon)^{1/4}$ and time scale $T_K = (\nu/\varepsilon)^{1/2}$, where $\nu$ is the kinematic viscosity of the fluid and $\varepsilon$ is the viscous dissipation rate of kinetic energy. Kolmogorov and Batchelor-Obukhov-Corrsin universal similarity of low order statistical parameters like spectra of velocity and temperature have been observed in all turbulent flows tested, Gibson 1991.

The inertial-vortex force per unit mass $\mathbf{v} \times \omega$ is the nonlinear force that causes turbulence and the turbulence cascade, where $\mathbf{v}$ is the velocity and $\omega$ is the vorticity. It appears in the conservation of momentum equation written in the form

$$\frac{\partial \mathbf{v}}{\partial t} = -\nabla B + \mathbf{v} \times \omega + \mathbf{F}_b + \mathbf{F}_C + \mathbf{F}_\nu + \cdots \quad (1)$$

where B is the Bernoulli group of mechanical energy terms $p/\rho + v^2/2 + gz$, p is the pressure, $\rho$ is the density, g is the acceleration of gravity, z is the distance above a reference level, $\mathbf{F}_b$ is the buoyancy force, $\mathbf{F}_C$ is the Coriolis force, $\mathbf{F}_\nu$ is the viscous force, and other forces are neglected. The Froude, Rossby and Reynolds numbers Fr, Ro, and Re are ratios of $\mathbf{v} \times \omega$ to $\mathbf{F}_b$, $\mathbf{F}_C$, and $\mathbf{F}_\nu$, respectively. Turbulence arises when $\nabla B$ is negligible and Fr, Ro, Re and all other such dimensionless ratios are greater than critical values, giving

$$\frac{\partial \mathbf{v}}{\partial t} = \mathbf{v} \times \omega \quad (2)$$

from (1). Fluid particles are accelerated in a direction perpendicular to both the velocity and vorticity, which is in opposite directions on opposite sides of a shear layer. Because $\mathbf{v}$ is zero at the shear layer and $\omega$ is zero far away, a maximum occurs in both the magnitude of $\mathbf{v} \times \omega$ and the local Reynolds number at distances $\sim L_K$ above and below any free shear layer of age > $T_K$, causing perturbations to be amplified by $\mathbf{v} \times \omega$ forces, as shown in Fig. 1.

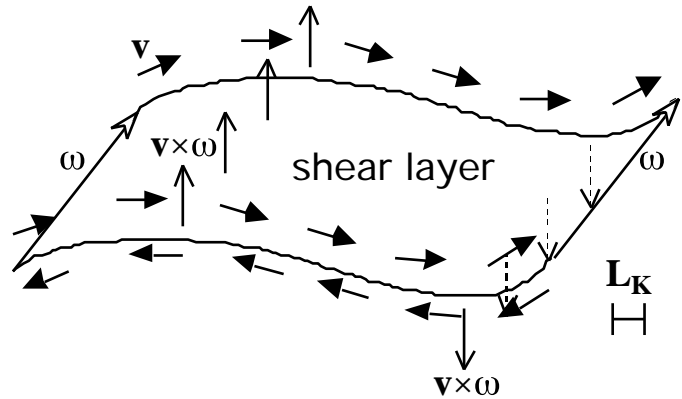

**Figure 1.** Instability of a shear layer to inertial-vortex forces $\mathbf{v}\times\omega$. Turbulent eddies form with Kolmogorov time scale $T_K$ at the Kolmogorov length scale $L_K$.

We see from Fig. 1 that the response of any shear layer to inertial-vortex forces is to roll up into an eddy with diameter $\sim L_K$ in a time $> T_K$. This is the basic turbulence mechanism. Turbulence eddies begin to form first at Kolmogorov length and time scales, the smallest possible, unless the flow is constrained by walls or forces larger than $\mathbf{v}\times\omega$.

Therefore, *turbulence is defined as an eddy-like state of fluid motion where the inertial-vortex forces of the eddies are larger than any other forces that tend to damp the eddies out*. All processes that produce shear layers produce turbulence by the mechanism of Fig. 1. Jets, wakes, boundary layers, and mixing layers are examples where shear layers are produced at solid surfaces, but shear layers can also form and become turbulent without solid boundaries as in the interior of large amplitude internal or surface waves or on tilted density interfaces in gravitational fields.

### 3. Direction of the turbulence cascade

From the definition, turbulence cascades from small scales to large. Eddies form at the Kolmogorov scale (Fig. 1), pair with neighboring eddies, and these pairs pair with neighboring pairs, etc., as shown in Figure 2. At each stage of the cascade, the average inertial-vortex forces are larger than any other force affecting the eddies, and these average inertial-vortex forces produce the next larger set of eddies by the same eddy pairing mechanism which formed them. Flow over a barrier is given as a generic representation of the wide variety of free shear layer sources that produce turbulence.

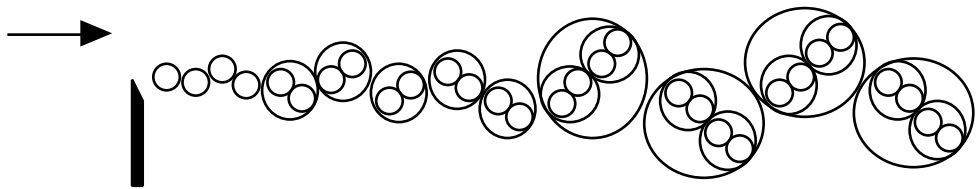

**Figure 2.** The turbulence cascade is always from small scales to large. Turbulence induces a random flow of external fluid and kinetic energy from large scales to small, but this is a nonturbulent cascade because irrotational flows have zero **v**×ω forces.

In Fig. 2, the volume of turbulent fluid increases downstream because external fluid with different velocities is entrained to supply the turbulent kinetic energy. Because the external fluid is irrotational, its flow is nonturbulent by definition. Inertial vortex forces in irrotational flows are zero. A nonturbulent energy cascade is induced in the external fluid from large scales to small that may be confused with the turbulence cascade because the external flow is randomized by its turbulent boundary conditions. This is the basis of the commonly accepted misconception that turbulence cascades from large scales to small, and that the cascade of turbulence from small scales to large clearly shown in Fig. 2 represents an inverse cascade that never occurs in three dimensional turbulence. The terminology "inverse cascade" is therefore inappropriate when applied to any turbulent flow since turbulence always cascades from small scales to large.

## 4. Definition of fossil turbulence

Fossil turbulence is defined as any fluctuation in a hydrophysical field produced by turbulence that is no longer turbulent at the scale of the fluctuation. Familiar examples of fossil turbulence include airplane contrails in the stratified atmosphere, skywriting, bubble patches from breaking surface waves, and milk patches in weakly stirred coffee. The time required for a turbulent patch to form by the processes of Figs. 1 and 2 is generally much less than the time required for particle diffusion processes to erase the scars in the numerous hydrophysical fields possibly affected by the turbulence. Buoyancy forces at their larger scales fossilize most patches of temperature and salinity microstructure detected in the ocean interior. Patches of clear air turbulence are generally fossilized at the scale of the aircraft well before they are encountered.

Fossil turbulence remnants preserve information about the hydrodynamic state of the fluid and flow that existed previously that may be difficult or impossible to recover otherwise, Gibson 1999. One of the oldest hydrophysical fossils is the cosmic microwave background radiation (CMBR), which are photons that decoupled at time $t = 10^{13}$ seconds after the big bang when the cooling plasma of hydrogen and helium turned to gas, redshifted by a factor $z = 1100$ to the microwave range observed from the white-hot 3000 K temperature of transition.

**5. The first turbulence and the first fossil turbulence**

Prior to the availability of space telescopes to measure the CMBR precisely it was assumed that the primordial plasma should be strongly turbulent. Fluctuations of density caused by cosmological turbulence were assumed to trigger galaxy formation due to gravitational instabilities, Ozernoy and Chibisov 1976. However, strong turbulence should produce temperature fluctuation values T/T in the range $10^{-1}$ to $10^{-2}$ rather than the small amplitude range $10^{-4}$ to $10^{-5}$ that was observed by the 1989 Cosmic Background Explorer satellite COBE and subsequent measurements. Cosmologists have inferred from this that rather than providing the source for turbulent kinetic energy production as previously assumed, the expansion of the universe somehow inhibits turbulence formation by an unknown mechanism. An alternative interpretation is that these observations prove that strong turbulence in the primordial plasma must have been inhibited by a combination of both buoyancy forces of gravitational structure formation and strong radiation-viscous forces, Gibson 1996. Radiation energy $E/c^2$ dominated matter for $t < 10^{11}$ seconds after the big bang. Photon, neutrino, and gluon viscosities very likely produced subcritical Reynolds numbers for times $10^{11} > t > 10^{-33}$ s after the strong force freeze out permitted such particles to exist, Gibson 2000. However, for $10^{11} < t < 10^{13}$ s, gluon and neutrino viscosities are irrelevant and photon viscosity is inadequate to inhibit either turbulence or gravity, suggesting this is the epoch of the first gravitational structure formation. Thus, the epoch of the first turbulence is either after $10^{11}$ s when radiation viscosity is negligible, or before $10^{-33}$ s when the quark-gluon plasma emerged from the inflation epoch. Was the universe turbulent in the epoch before inflation and after the Planck time $T_P = 10^{-43} < t < 10^{-35}$ s? It appears that not only was the initial flow powerfully turbulent, but the turbulence process of the big bang created the universe.

Physicists describe small-scale structure using quantum mechanics and large-scale structure using Einstein's equations of general relativity. Both of these theories break down at the initial instant and singularity of the big bang, producing infinite energy integrals and probabilities greater than one, Peacock 2000. Gravity is not included in the tensor and scalar fields of standard wave-particle physics theory, and the fundamental particles of nature (fermions, vector and Higgs bosons) and the probability waves of quantum field theory do not emerge from Einstein's theory of space-time-gravity. Super string-M-theory is the closest approach to a synthesis currently available, Greene 1999. No theory of quantum-gravitational-dynamics (QGD) is accepted that provides a quantitative, or even qualitative, description of the first stages of the universe. Linear perturbation theories fail. The Einstein assumption of ideal fluids becomes untenable. Powerful clues constrain quantum mechanical grand unified theories (GUT) at the very high temperatures and very small scales that can be inferred by dimensional analysis based on light speed c, Plank's constant h, Newton's constant G, and Boltzmann's constant k. This is the domain of Planck scales and strong nonlinearity. Remarkably, the concept of turbulence beginning at Planck scales has been overlooked in cosmological modeling.

According to the standard model of the hot big-bang universe, space-time-energy emerged from the vacuum spontaneously at Planck scales because of the slight possibility of such an event allowed by Heisenberg's uncertainty principle, which states that the uncertainty of the energy of a particle E multiplied by the uncertainty of its time of existence t is a constant h, termed the Planck constant. The Planck mass $m_P = (ch/G)^{1/2}$ is found where the Compton (de Broglie) wavelength $L_C = h/mc$ of an object with mass m equals the Schwarzschild radius $L_S = Gm/c^2$ of a black hole, where c is the speed of light and G is Newton's gravitational constant. The Planck entropy $S_P$ is equal to the Boltzmann constant k, and gives a minimum black hole specific entropy $s_P = S_P/m_P$, maximum black hole temperature $T_P = (c^5h/Gk^2)^{1/2}$, and minimum black hole (Hawking) evaporation time $t_P = (hG/c^5)^{1/2}$, Greene 1999. Grand unified theories suggest all non-gravitational forces of nature (weak, strong, electromagnetic) are equivalent and converge at $10^{-4}$ times the Planck temperature, or $10^{28}$ K. Superstring-M-theory suggests the force of gravity also converges at this temperature, Greene 1999 Fig. 14.2. This simplicity of Planck scale conditions permits high Reynolds number turbulence and turbulent mixing. Thus, the quantum gravitational dynamics (QGD) time period is the epoch of big-bang turbulence.

Figure 3 gives the Heisenberg, Compton and Schwarzschild equations, the fundamental constants c, h, G, k, the Planck m, L, t, and T scales, and the Planck parameters of the first turbulence. Substituting the Planck viscosity $\nu_P = c^2 t_P$ and Planck dissipation rate $\varepsilon_P$ (Fig. 3) into the definitions of the Kolmogorov length and time scales give $L_K = L_P$ and $T_K = t_P$. The Planck thermal diffusivity $\alpha_P = c^2 t_P$, so the Batchelor scale $L_B = L_K(\alpha/\nu)^{-1/2}$ also equals $L_P$, where $\nu/\alpha$ is the Prandtl number $Pr_P = 1$. The Planck mass, length, time, and temperature scales can be found by dimensional analysis from the fundamental constants. Planck energy, power, viscous dissipation rate, specific entropy, density and gravitational force scales of Fig. 3 appear from either dimensional analysis or by combinations of $m_P$, $L_P$, $t_P$, and $T_P$. For example, the Planck inertial-vortex force per unit mass $[\mathbf{v} \times \omega]_P = L_P/t_P^2$ equals the Planck gravitational force per unit mass $g_P = F_{GP}/m_P = [c^7/hG]^{1/2} = 5.7 \times 10^{51}$ m s$^{-2}$ (Fig. 3 bottom line). This fundamental nonlinear force of the turbulence mechanism arises to drive the big bang when a Planck particle and a Planck antiparticle interact to form their first quantum state, which is a spinning pair forming an extreme Kerr black hole termed a Planck-Kerr particle. Electron-positron pair production and positronium formation is an analogous process known to occur at lower temperatures in supernova. The Schwarzschild metric of nonspinning black holes has been known since 1916, and the Kerr metric of spinning black holes has been known since 1963, Peacock 2000 (p 51-60). Spinning Kerr black holes are capable of Hawking radiation (pair production at the event horizon) with 42% efficiency for prograde collisions, Peacock 2000 (p 61). This produces the inertial-vortex force and the transfer of angular momentum and energy to larger scales that drives big bang turbulence.

$$h = \Delta E \times \Delta t = E_P t_P$$
$$L_{Compton} = h/mc = L_{Schwarzschild} = Gm/c^2$$
$$\text{when } m = m_P = [ch/G]^{1/2}$$

$c = 3 \times 10^8$ m s$^{-1}$     $m_P = [ch/G]^{1/2} = 2.12 \times 10^{-8}$ kg
$h = 1.05 \times 10^{-34}$ kg m$^2$ s$^{-1}$     $L_P = [hG/c^3]^{1/2} = 1.62 \times 10^{-35}$ m
$G = 6.67 \times 10^{-11}$ m$^3$ kg$^{-1}$ s$^{-2}$     $t_P = [hG/c^5]^{1/2} = 5.41 \times 10^{-44}$ s
$k = 1.38 \times 10^{-23}$ kg m$^2$ s$^{-2}$ K$^{-1}$     $T_P = [c^5h/Gk^2]^{1/2} = 1.40 \times 10^{32}$ K

$$E_P = [c^5h/G]^{1/2} = 1.94 \times 10^9 \text{ kg m}^2 \text{ s}^{-2}$$
$$P_P = c^5/G = 3.64 \times 10^{52} \text{ kg m}^2 \text{ s}^{-3}$$
$$\varepsilon_P = [c^9/hG]^{1/2} = 1.72 \times 10^{60} \text{ m}^2 \text{ s}^{-3}$$
$$s_P = [k^2G/ch]^{1/2} = 6.35 \times 10^{-16} \text{ m}^2 \text{ s}^{-2} \text{ K}^{-1}$$
$$\rho_P = c^5/hG^2 = 5.4 \times 10^{96} \text{ kg m}^{-3}$$
$$F_{GP} = Gm_P^2/L_P^2 = 1.1 \times 10^{44} \text{ kg m s}^{-2}$$
$$[\vec{v} \times \vec{\omega}]_P = g_P = [c^7/hG]^{1/2} = 5.7 \times 10^{51} \text{ m s}^{-2}$$

**Figure 3.** Planck scales and Planck parameters of the first turbulence.

Once the first Planck particle, Planck antiparticle, and Planck-Kerr particle emerge, their enormous temperatures increase the probability of more Planck particle pairs forming nearby, causing the big bang turbulent cascade to larger and larger scales that formed the universe in our present model. Random interactions between Planck particles and Planck-Kerrs caused larger scale aggregations of Planck pairs with the same rotational sense, just like the turbulence cascade of Fig. 2. Each pair production event produces more space-time and momentum-energy near the Planck temperature of $10^{32}$ K. No mechanism exists to stop this cascade equivalent to the gluon viscosity and photon viscosity that inhibits turbulence at lower temperatures. Only when temperatures cool to the strong force freeze out temperature of $10^{28}$ K can gluons and quarks exist, so that long range momentum transfer by gluon radiation can produce viscous damping of turbulence and inflation by the negative stresses of bulk viscosity. Entropy production resulting from this big bang turbulent mixing of velocity and temperature fluctuations in the initial expanding universe are equivalent to entropy production from viscous and temperature dissipation rates and in less exotic turbulent mixing flows, so that the temperature spectrum $\phi_T$ should depend only on these rates and the wavenumber k, giving $\phi_T = \beta \chi \epsilon^{-1/3} k^{-5/3}$ by dimensional analysis with $\beta$ a universal constant of order 1 as first shown by Corrsin and Obukhov. Figure 4 illustrates the mechanism. From Fig. 3 we see that the total entropy production of the hot big bang turbulence event was only about 10 J/K, a remarkably small amount equivalent to boiling a milligram of water. Most of the entropy of the universe was produced after inflation, when the free quarks of the quark gluon plasma condensed to form hadrons (protons, mesons) and the hadrons formed the light elements at the much lower temperatures and later times of the quantum-chromo-dynamics (QCD) and quantum-electro-dynamics (QED) epochs.

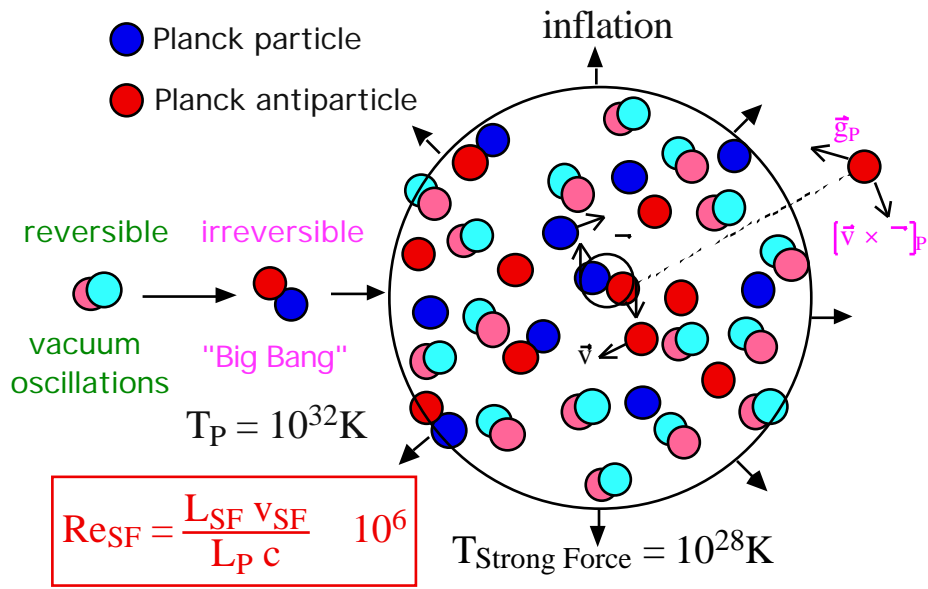

**Figure 4.** Physical process of the first turbulence. Reversible vacuum oscillations at Planck scales have a finite probability of forming a Planck particle-antiparticle Planck-Kerr pair that can trigger irreversibilities and the formation of the universe. Quantized rotation states (extreme Kerr black holes, center) can form and produce gravitational inertial-vortex forces (upper right) that diffuse the spin, slow the annihilation, and homogenize the velocity and temperature fluctuations by turbulent eddy formation and mixing until the strong force freeze-out temperature is reached.

Exponential inflation of space caused by negative viscous and false vacuum pressures then fossilizes the turbulent temperature field.

Because the temperatures of the quantum gravitational dynamics (QGD) epoch were too high for smaller particles to exist, momentum transport was only by short range Planck particle interactions, giving small kinematic viscosities $\nu_P \approx cL_P$ and large Reynolds numbers $Re = vL/cL_P$ at scale L with velocity $v \approx (2kT)^{1/2}$. According to inflation theory, Guth 1997, the big bang cascade terminates when temperatures decrease to the strong force freeze-out temperature $T_{SF} = 10^{28}$ K at time $10^{-35}$ seconds where quarks and gluons form, with $L = L_{SF} = 10^8 L_P$, $v_{SF} = 10^{-2}$ c, and maximum Reynolds number or order $10^6$ as shown in Fig. 4. The only heavy particle stable in the temperature range $T_P > T > T_{SF}$ is the magnetic monopole with mass about $10^{-2}$ $m_P$. Guth found that the lack of observed magnetic monopoles at present implies an inflationary epoch between time $t = 10^{-35}$ and $10^{-33}$ s, driven by the equivalent of antigravity forces from a false vacuum negative pressure, according to general relativity theory and Einstein's equations which relate space time geometry to the energy momentum tensor. During inflation (by a factor of about $10^{25}$) the Planck scale fluctuations at $10^{-35}$ m were stretched to $10^{-10}$ m, the strong force scale fluctuations at $ct_{SF} = 3 \times 10^{-27}$ m were stretched to $3 \times 10^{-2}$ m, and geometry was flattened to a nearly Euclidian state at large scales. All the temperature fluctuations of the first turbulence were stretched by this exponential inflation of space to scales much larger than the horizon scale $L_H = ct$ of $3 \times 10^{-25}$ m existing at the end of the inflation epoch $10^{-33}$ s, thus producing the first fossil temperature turbulence. The expected spectrum for high Reynolds number turbulent temperature fluctuations is the universal Corrsin-Obukhov form $\Phi_T = \beta \varepsilon^{-1/3} k^{-5/3}$, where k for the first turbulence is in the range $1/L_{HSF} > k > 1/L_P = 1/L_B$.

After inflation, the fossil-temperature-turbulence (fossil-curvature-turbulence and fossil-vacuum-turbulence) fluctuations continue to stretch as the universe expands, from the general relativity theory of Einstein. Fossil Planck scale fluctuations reentered the horizon first because they have the smallest scales, but not until the universe cooled to the electoweak freeze-out temperature so that radiation in the form of neutrinos and photons with mean free paths much larger than the electron separations provided viscosities sufficient to damp out any turbulence, Gibson 2000. Because nuclear reactions are very sensitive to temperature, the temperature fossils of the first turbulence imprinted the mass density and species concentrations of light elements formed by nucleosynthesis when the first protons, electrons, neutrons and alpha particles were formed before $t = 10^2$ s, Weinberg 1977, converting the temperature fossil to density and concentration fossils. Further cooling decreased the neutrino-electron scattering cross section to negligible levels. Thus the universe became completely transparent to all neutrino species after about $10^2$ s, Weinberg 1972, producing a cosmic neutrino radiation background picture of the universe at this neutrino decoupling time that may some day be observed when high resolution neutrino telescopes are invented.

Photon viscosity values produced viscous-gravitational scales larger than the horizon scale $L_H$ until just before the time of first structure at $10^{12}$ s, Gibson 1996, when the proto-supercluster mass of $10^{46}$ kg just matched the horizon mass $L_H^3$, where is the density computed from general relativity theory, Weinberg 1972. Diffusion of the neutrino-like nonbaryonic dark matter filled the proto-superclustervoids, proto-clustervoids, and proto-galaxyvoids formed in the plasma epoch, reducing gravitational driving forces and arresting further structure formation until after the plasma-gas transition time $10^{13}$ s (300,000 years) when the temperature was about 3000 K, Gibson 2000. At that time atoms formed and photons decoupled from electrons, giving the cosmic microwave background (CMB) image of the universe that we observe today, redshifted by the expansion of the universe into the microwave bandwidth.

In the big bang turbulence model presented above, an enormous amount of formalism regarding general relativity theory, quantum field theory, particle physics, and M-superstring theory have been passed over in order to focus on the qualitative mechanisms of nonlinear viscous flows at Planck scales. The mathematical issues of introducing the extra dimensions of string theories to the Einstein field equations for an imperfect fluid are discussed by Paul 2001, who predicts a power law inflation of three dimensional physical space can be driven by viscous stresses. A discussion of relativistic radiation viscosity is given by Chen and Spiegel 2001. They show that the bulk viscosity of an expanding flow, such as that of the expanding universe, cannot be ignored even for the classical case of rigid spherical particles, and can be much larger than the shear viscosity or thermal diffusivity.

## 6. Discussion

High resolution maps and spectra of Cosmic Microwave Background (CMB) temperature fluctuations have recently been obtained from telescopes carried by balloons to altitudes above atmospheric interference, extending the 1989 Cosmic Background Experiment (COBE) space telescope observations to smaller scales and higher precision. Hu 2000 compares a collection of spectral points to various cosmological models, and the present turbulence model is compared to these in Figure 5.

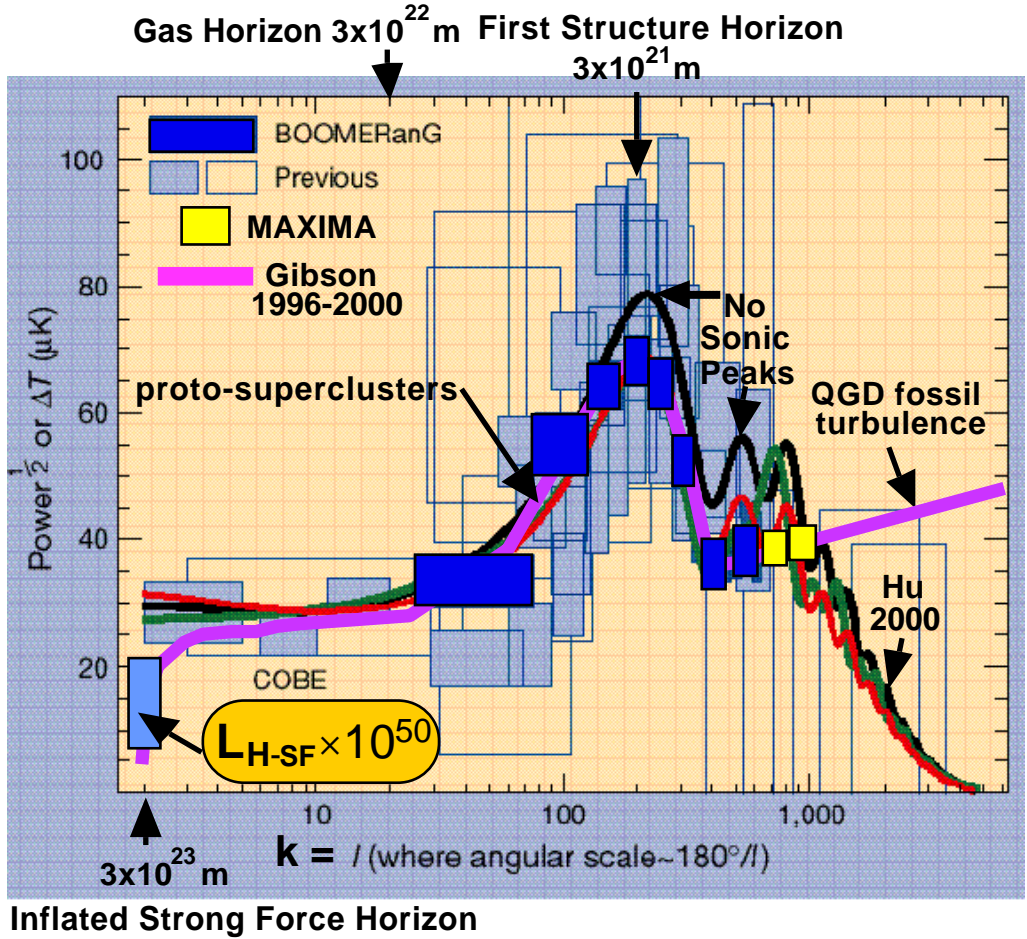

**Figure 5.** Cosmic microwave background temperature spectra (k = l) compared to standard cosmological models (from the Figure in Hu 2000) and the Gibson 1996-2000 predictions of early structure and QGD fossil turbulence $\sim k^{1/6}$. The largest scale (all sky) COBE point (missing from the Hu 2000 Figure) shows a significant decrease as expected for the strong force horizon scale $L_{H\text{-}SF}$, fossilized and inflated by a factor of $10^{50}$. The more precise BOOMERanG and MAXIMA points are in poor agreement with sonic peaks plus Tilted forms assumed by the standard models. The peak at wavelength $3 \times 10^{21}$ m thus represents the first structure formation (proto-superclustervoids), formed at smaller scales and earlier times than the gas-plasma horizon wavelength $= 2 \pi L_H \approx 10ct = 3 \times 10^{22}$ meters. The expected fossil Planck scale peak is off-scale to the right at $\lambda = 3 \times 10^{15}$ m and $\Delta T = 430$ µK (see Fig. 6).

As shown in Fig. 5, the theoretical predictions of standard cosmological models and the present theory strongly diverge at high wavenumbers. The Harrison-Zel'dovich $k^2$ dissipation spectral form is flat. Tilted forms have decreasing tails (Hu 2000, Fig. 5) for large k and large Legendre multipole index l, and further droop results by assuming diffusive damping of temperature at small scales. In Figure 6 (dashed box) the bump marked BOOMERANG represents the prominent peak of Fig. 5. The peak occurs at a wavelength $= ct/3^{1/2}$ corresponding to the sound speed of the plasma epoch, where c is the speed of light, which has been taken by several authors to be evidence that the peak represents sonic oscillations of the plasma. However, Gibson 2000 argues that any plasma sonic oscillations would be strongly damped by photon-viscous forces due to Thomson scattering of photons with electrons that are strongly coupled to ions by electric forces, and that the spectral peak of Fig. 5 represents an earlier, smaller, horizon scale $L_{H\text{-structure}}$ of proto-galaxy-supercluster-void structures of the first gravitational instability. No cosmological sound source exists for such a strong sonic peak, with $T/T$ $10^{-4}$. Such fluctuations are too small to be consistent with fully developed turbulence in the plasma, but are too large to be caused by sonic fluctuations in this super-viscous fluid with no sonic sources. No diffusive cutoff is expected at high frequencies because the fluctuations should reflect fossil-density-turbulence created by nucleosynthesis triggered by big bang turbulence remnants, and these density fluctuations have smaller diffusivities than temperature, which is transferred by photon radiation. Slight drooping of large k spectral estimates may exist from undersampling errors due to intermittency, since only a small fraction of the sky has been sampled by the present measurements and strong intermittency can be expected in the small scales of such high Reynolds number fossil turbulence.

The suggested interpretations of the observations of Fig. 5 are shown schematically in Fig. 6, where spectral forms are extended by 5 decades to the fossilized turbulent Planck scale. If a QGD fossil turbulence spectrum were used with the sonic standard model predictions instead of the Tilted forms, the higher order sonic peaks would become even more inconsistent with the data and the interpretation of the primary peak as a sonic oscillation even less tenable. Higher resolution, higher precision CMB observations planned from balloons and new space telescopes may resolve the contradictory predictions in Figs. 5 and 6. If the peak is not due to sonic oscillations, then the Gibson 1996 suggestion that it reflects the first gravitational formation of structure in the plasma epoch is supported, contrary to the Jeans 1902 sonic criterion for gravitational structure formation.

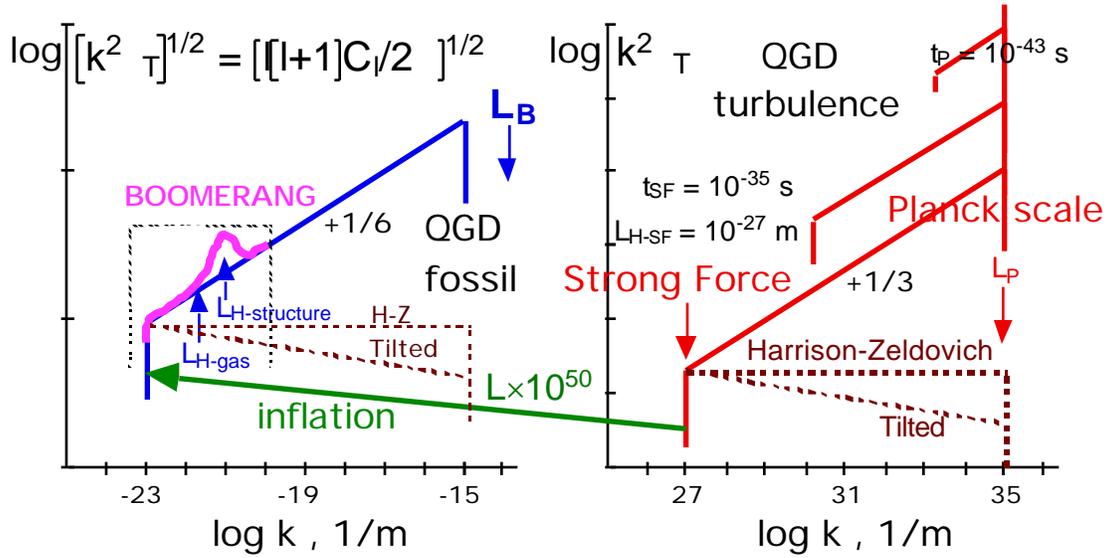

**Figure 6.** Dissipation spectra for QGD temperature fluctuations. On the right, the present theory predicts turbulent mixing spectral forms that peak at the smallest scale, compared to flat Harrison-Zel'dovich and Tilted forms that peak at the largest scale. A schematic version of the fossilized spectra is shown on the left, after inflation by $10^{50}$ as indicated by observations in Fig. 5 (box). All fluctuations are larger than the Batchelor scale of diffusive damping $L_B$ of about $10^{14}$ m for ions in the plasma.

The Jeans criterion follows from a linear perturbation stability analysis of the inviscid Euler equations with gravity, reducing the stability problem to one of gravitational acoustics. By the Jeans theory, sound waves are unstable in a gas of density if the time $L/V_S$ required to propagate a wavelength L at sound speed $V_S$ is greater than the time required for gravitational freefall $(G)^{-1/2}$. Density fluctuations smaller than the Jeans scale $L_J = V_S(G)^{-1/2}$ are assumed to be gravitationally stable in standard cosmological models, where G is Newton's gravitational constant (Fig. 3). However, Gibson 1996 and Gibson 2000 suggest that viscous, turbulent, and other forces may also set length scale criteria for gravitational structure formation smaller or larger than $L_J$. Because the horizon scale $L_H$ in the plasma epoch is always smaller than $L_J$, standard cosmological models based on Jeans' theory such as Weinberg 1972, Silk 1989, Kolb and Turner 1990, Peebles 1993, Padmanabhan 1993 and Rees 2000 assume no gravitational structures can form. The Jeans criterion is invalid during the plasma epoch, and is almost completely irrelevant in the gas epoch except for the formation of globular star clusters. Viscous and turbulence forces are not able to prevent gravitational structure formation when either the viscous Schwarz scale $L_{SV}$ or the turbulent Schwarz scale $L_{ST}$ become smaller than $L_H$, where $L_{SV} = (/G)^{1/2}$, $L_{ST} = {}^{1/2}/(G)^{3/4}$, is the rate of strain of the gas with density and kinematic viscosity ,

and   is the viscous dissipation rate, Gibson 1996. As mentioned previously, this event occurred at about t = $10^{12}$ s, or 30,000 years, so that supercluster masses were the first gravitational structures to form rather than the last as assumed in standard cosmological models. Therefore, the spectral peak of Fig. 5 is due to gravitational structure formation nucleated by the fossil density fluctuations from the first turbulence at viscous-gravitational and turbulent-gravitational scales in the plasma as soon as these Schwarz scales became smaller than $L_H$, Gibson 2000.

## 7. Conclusions

The chaotic quantum gravitational dynamics (QGD) epoch at Planck scales in the beginning of the universe (Section 5) satisfies the definition of turbulence (Section 2) and the consequent direction of the turbulent cascade (Section 3) from small scales to large. Inertial-vortex forces that define and drive ordinary turbulence are identified with an efficient Hawking radiation of energy and angular momentum that results when nonspinning, Planck-scale, Schwarzschild black holes interact with Planck-scale, spinning, Planck-Kerr black holes. Because this gravitational-centrifugal Planck-scale inertial-vortex force matches the enormously large Planck-scale gravitational force, it produces big bang turbulence and the universe, and represents a new, fundamental, fifth force of nature. By an extension of grand unified quantum mechanical theory the electromagnetic, weak, strong, gravitational, and inertial-vortex forces are all equivalent in the QCD temperature range $10^{28}$ < T < $10^{32}$ K, and only Planck particles, Planck antiparticles, Plank-Kerr spinors, and magnetic monopoles can exist. Viscous production of vorticity occurs at Planck scales, and this is convected and homogenized by turbulence to larger scales as the horizon expands and the Reynolds number increases. Turbulence homogenizes the increasing temperature and entropy fluctuations until the universe cools to the strong force freeze out temperature at which gluon-viscosity damps the turbulence and triggers inflation. Gluon bulk viscosity and the Guth false vacuum negative pressures at the strong force freeze out time exponentially inflate space to atomic dimensions; that is, from scales smaller than the strong force horizon at $10^{-27}$ m to inflated fossil Planck scales larger than $10^{-10}$ m, giving fossil temperature turbulence remnants much larger than the horizon scale of $10^{-24}$ m at the end of inflation.

Besides being the first turbulence, big bang turbulence was also one of the most powerful turbulence events in nature, despite its small size and short duration. Laboratory turbulence generally has at most one or two decades of inertial subrange and oceanic and atmospheric turbulence three or four, compared to eight decades for big bang turbulence.

The Planck power of $3.6 \times 10^{52}$ kg m$^2$ s$^{-3}$ (Fig. 3) driving the big bang event is $10^4$ greater that the radiated power of all the stars in all the galaxies within our present horizon.

It appears from the cosmic microwave background temperature spectrum of Figure 5 that remnants of the big bang process at the strong-force-freeze-out scale have been inflated by a factor of $10^{50}$ to scales ten times larger than the gas horizon wavelength of $3 \times 10^{22}$ existing at the time of the plasma to gas transition, and that something like big bang turbulence must have occurred before this inflation to explain the isotropy of the measured fluctuations, which are independent of direction on the sky. If the dominant spectral peak at $3 \times 10^{21}$ m wavelength has been stretched by a factor of $10^3$ corresponding to the gas redshift $z = 1100$ then its scale matches the observed $10^{24}$ m scale of supervoids and superclusters. This supports the interpretation that the spectral peak is due to the first gravitational structure formation, not sound, and that these galaxy cluster voids have simply expanded with the universe after fragmentation and that cluster densities have rarely increased by any collapse or accretion process, Gibson 2000. Some evidence of these supercluster complexes has been reported on the $10^{25}$ m scales of the stretched strong-force freezeout scales of the big bang as they appear within the present horizon of $10^{26}$ m. The most distant galaxies have measured redshifts $z > 3$ and are already strongly clustered, contrary to models where clustering occurs at the last rather than the first stages of gravitational structure formation.

Thus, Kolmogorov universal similarity of the velocity fluctuations and Corrsin-Obukhov-Batchelor universal similarity of the temperature fluctuations in the expanding, cooling, hot big-bang turbulence universe are indicated, with maximum amplitude of the temperature gradient spectrum occurring at the small Planck length scale source of the temperature fluctuations before fossilization. Inflation of space at the strong-force-freeze-out temperature fossilized eight decades of turbulent temperature fluctuations by stretching beyond the horizon scale of causal connection, and these provided chaotic seeds for nucleosynthesis and gravitational structure formation after they reentered the horizon. Baryonic (ordinary) matter concentrations preserved the fossil temperature turbulence patterns formed in the QGD epoch as density and light element concentration fluctuations, which reappear as temperature fluctuations in the cosmic microwave background (CMB) by the gravitational Sachs-Wolfe red shift mechanism.

The present paper is dedicated to the memories of George Batchelor and Chuck Van Atta, and was presented on the occasion of John Lumley's 70$^{th}$ birthday celebration at Cornell University.